\documentclass[pra,twocolumn,groupedaddress,showpacs,floatfix]{revtex4}

\usepackage{graphics}
\usepackage{graphicx}
\usepackage{bm}
\usepackage{amsmath}
\usepackage{amsfonts}
\usepackage{amssymb}
\usepackage{latexsym}
\usepackage{color}

\begin{document}

\title{Laser cooling of a trapped particle with increased Rabi frequencies}
\author{Tony Blake, Andreas Kurcz, Norah S. Saleem, and Almut Beige}
\affiliation{The School of Physics and Astronomy, University of Leeds, Leeds, LS2 9JT, United Kingdom} 

\date{\today}

\begin{abstract}
This paper analyses the cooling of a single particle in a harmonic trap with red-detuned laser light with fewer approximations than previously done in the literature. We avoid the adiabatic elimination of the excited atomic state but are still interested in Lamb-Dicke parameters $\eta \ll 1$. Our results show that the Rabi frequency of the cooling laser can be chosen higher than previously assumed, thereby increasing the effective cooling rate but {\em not} affecting the final outcome of the cooling process. Since laser cooling is already a well established experimental technique, the main aim of this paper is to present a model which can be extended to more complex scenarios, like cavity-mediated laser cooling. 
\end{abstract}


\maketitle

\section{Introduction} \label{Intro}

The idea of laser cooling for neutral atoms was first suggested by H\"ansch and Schalow \cite{Haensch} and independently for trapped ions by Wineland and Dehmelt \cite{WinelandDehmelt}. Since light carries momentum as well as energy, scattering light on matter can result in significant changes of the vibrational energy of massive particles. Over the last decades, laser cooling techniques have been designed which allow the cooling of atoms and ions to the the micro and nanokelvin temperatures needed for quantum coherence and degeneracy \cite{PhilipsNL,ChuNL,Cohen}. Examples of which are Sisyphus cooling \cite{CohenPhilips} and evaporative cooling \cite{Ketterle}. In this paper we focus our attention on laser sideband cooling of a particle in a harmonic potential \cite{sideband} which can be used for example to transfer single ions to their motional ground state with a very high precision \cite{wineland2,more2}. 

The idea of laser sideband cooling of a single trapped particle is relatively straightforward. If the trapping potential of the particle is strong enough for its motion to become quantised and if the corresponding phonon frequency is much larger than the spontaneous decay rate of the excited electronic state, it becomes possible to resolve the motional sidebands with the cooling laser. As a consequence, the excitation of the excited electronic state of the trapped particle on the red sideband is most likely accompanied by the annihilation of a phonon. When followed by the spontaneous emission of a photon, the particle returns into its electronic  ground state without regaining a phonon which implies cooling. The cooling cycle only stops when the particle reaches a state with almost no phonons. The final population of excited vibrational states is in general very small and only due to highly off-resonant excitations of the ground state of the atom-phonon system.

The theory of laser cooling has already been studied in great detail in the literature. For reviews on this topic see for example  Refs.~\cite{Stenholm2,key,RevMod,ions}. The first theoretical discussion of laser cooling with red-detuned light based on a combination of simple classical and quantum ideas can be found in Ref.~\cite{sideband} by Wineland and Itano. Lindberg and Stenholm later introduced the tool for a full quantum treatment of laser cooling by deriving a master equation for spontaneously emitting atoms with atomic recoil included \cite{Stenholm} (cf.~also~Refs.~\cite{Sten,Lindberg,Stenholm4,Stenholm3,dalibard}).
An alternative but consistent analysis of the laser cooling of trapped ions in a running and in a standing wave configuration has been presented by Cirac {\em et al.} in Ref. \cite{sideband2}. The main result of these papers is a cooling equation of the form
\begin{eqnarray}
\dot m &=& - (A_- - A_+) \, m + A_+ \, ,
\end{eqnarray}
where $m$ denotes the mean phonon number. The $A_\pm$ can be interpreted as transition rates between states with different phonon numbers and relate to actual cooling and heating rates. Moreover, one can show that the stationary state phonon rate $m_{\rm ss}$ is given by \cite{Stenholm,Stenholm2,sideband2,ions}
\begin{eqnarray} \label{apm}
m_{\rm ss} &=& {A_+  \over A_- - A_+} \, . 
\end{eqnarray}
Experimental results confirm the general dependence of this stationary state phonon number on the emission rate of the excited electronic state of the trapped particle $\Gamma$ and on its phonon frequency $\nu$ \cite{more}.

The purpose of this paper is to analyse the cooling process of a single trapped particle in a harmonic potential with fewer approximations than previously done in the literature. For example, we avoid the adiabatic elimination of the excited atomic state, thereby obtaining an analysis of the cooling process which applies for a much wider range of Rabi frequencies $\Omega$ of the cooling laser. As a result we find that $\Omega$ does not have to be small compared to the spontaneous decay rate $\Gamma$ and the laser detuning $\Delta$. For example, in the weak confinement regime with $\nu \ll \Gamma$, the Rabi frequency $\Omega $ can be as large as $0.3 \, \Gamma$ and in the strong confinement regime with $\Gamma \ll \nu$, the Rabi frequency $\Omega $ can be as large as $0.3 \, \nu$ without noticeably increasing the stationary state phonon number $m_{\rm ss}$. This is an interesting observation, since the effective cooling rate $\gamma_{\rm c}$ scales as $\Omega^2$ which increases rapidly when $\Omega$ increases.

Our calculations are more straightforward than previous calculations, since we replace the atomic raising operator $\sigma^+$ and the phonon annihilation operator $b$ by two new operators $x$ and $y$. These correspond to particles that are neither atoms nor phonons and commute with each other. Most importantly, they provide a representation of the Hamiltonian which no longer contains atomic displacement operators. Instead it depends on terms like $x^\dagger x (y-y^\dagger)$ which take non-linear effects in the atom-phonon interaction into account \cite{Vogel}. In the following we use this Hamiltonian to obtain a manageable set of cooling equations which are differential equations for the time derivatives of expectation values. As in Refs.~\cite{Stenholm,Stenholm2,sideband2,Stenholm4,ions}, we are interested in the dynamics of the cooling process on the very slow time scale given by the cooling rate $\gamma_{\rm c}$ which scales as $\eta^2$ with $\eta \ll 1$. The only approximation involved in the following calculations of stationary state phonon numbers and effective cooling rates is to neglect higher order terms in the Lamb-Dicke parameter $\eta$.

The main purpose of this paper is to establish and test a framework for the modelling of laser cooling which can be extended relatively easily to more complex cooling scenarios like cavity-mediated laser cooling \cite{cavity} and the study of possible quantum optical heating mechanisms in sonoluminescence experiments \cite{SL}. Cavity-mediated laser cooling for example has many similarities with laser cooling but due to being more complex there is more flexibility in choosing different scenarios and experimental parameters \cite{cavity2}. 

There are six sections in this paper. Section \ref{model} introduces the master equation of a single laser-driven trapped particle. Section \ref{standard} uses this master equation to derive a closed set of 23 cooling equations. These simplify to a set of five equations in the weak confinement regime and to a single equation in the strong confinement regime, respectively. In Section \ref{stability} we show that the phonon coherences and the mean phonon number $m$ always reach their stationary state. Section \ref{process} presents cooling rates and stationary state phonon numbers and compares analytical and numerical results. Finally, we summarise our findings in Section \ref{conc}. Mathematical details are confined to Apps.~\ref{app}--\ref{appF}.

\section{Theoretical model} \label{model}

Let us start by introducing the theoretical model and the experimental setup. It consists of a single cooling laser which drives a strongly confined single particle. As long as the trapping potential is approximately harmonic, we can describe the motional states of the atom by a harmonic oscillator Hamiltonian. In this section, we consider the motion of the trapped particle as quantised and introduce the model which allows us to predict the time evolution of its mean phonon number $m$. The purpose of the cooling laser is to minimise the number of phonons in the motion of the particle in the direction of the laser. Cooling other vibrational modes requires additional cooling lasers.

\subsection{The Hamiltonian}

The Hamiltonian of a single trapped particle inside the free radiation field and with external laser driving can be written as 
\begin{eqnarray} \label{1.23} 
H &=& H_{\rm nuclei} + H_{\rm electron} +  H_{\rm field} + H_{\rm dip}  \, .
\end{eqnarray}
The first two terms are the free energy of the electronic states and of the quantised vibrational modes of the trapped particle. The third term describes the energy of the surrounding free radiation field. The last terms take the dipole interaction of the electronic states with the present electromagnetic fields, i.e.~the laser and the free radiation field, into account. We now have a closer look at every term in this equation.

Suppose the particle is effectively a two-level system with ground state $|0\rangle$ and excited state $|1\rangle$ and the energies $\hbar \omega_0$, $\hbar \nu$, and $\hbar \omega_k$ denote the energy of a single atomic excitation, of a single phonon excitation, and of a single excitation of the modes of the free radiation field, respectively.  Then we have
\begin{eqnarray} \label{1.23b}
H_{\rm electron} &=& \hbar \omega_0 \, \sigma^+ \sigma^- \, , \notag \\
H_{\rm nuclei} &=& \hbar \nu \, b^\dagger b \, , \notag \\
H_{\rm field} &=& \sum_{{\bf k} \lambda} \hbar \omega_k \, a_{{\bf k} \lambda}^\dagger a_{{\bf k} \lambda} \, , 
\end{eqnarray}
where the operators $\sigma^{-} \equiv |0\rangle \langle1|$ and $\sigma^{+}\equiv|1\rangle \langle0|$ are the atomic lowering and raising operator and where $b$ and $a_{{\bf k} \lambda}$ are the phonon annihilation operator and the annihilation operator of a photon with wavevector ${\bf k}$ and polarisation $\lambda$, respectively. These operators obey the commutator relation 
\begin{eqnarray} \label{comms}
\big[a_{{\bf k} \lambda},a_{{\bf k} \lambda}^\dagger \big] = \big[b,b^{\dagger} \big] &=& 1   
\end{eqnarray}
which is the usual commutator relation for bosonic annihilation operators. All other photon commutators are equal to zero.

The final term in Eq.~(\ref{1.23}) describes the dipole interaction of the electronic states $|0 \rangle$ and $|1 \rangle$ of the trapped particle with the free radiation field and the applied laser field. Within the usual dipole approximation \footnote{This means, we assume that the size of the atom is small compared to the relevant optical wavelength.}, it can be written as
\begin{eqnarray} \label{HL1}
H_{\rm dip} (t) &=&  e\textbf{D} \cdot \big[ \textbf{E}_{\rm field} (\textbf{R}) + \textbf{E}_{\rm L} ({\bf R},t) \big] \, .
\end{eqnarray}
Here $e$ is the electron charge, ${\bf D}$ is the dipole moment of the particle, i.e.~the position operator of its outer electron with respect to the atomic nuclei at position ${\bf R}$, while $\textbf{E}_{\rm field} ({\bf R})$ and $\textbf{E}_{\rm L} (\textbf{R},t)$ denote the electric field amplitude of the free radiation field and of the laser field at time $t$, respectively. The dipole moment ${\bf D}$ equals
\begin{eqnarray} \label{DDD}
{\bf D} &=& \textbf{D}_{01} \, \sigma^{-}+\mbox{H.c.} \, ,
\end{eqnarray}
where $\textbf{D}_{01}$ is a 3-dimensional complex vector. The electric field operators are given by
\begin{eqnarray}
\textbf{E}_{\rm field} (\textbf{R}) &=& {\rm i} \sum_{{\bf k} \lambda} \sqrt{{\hbar \omega_k \over 2 \epsilon_0 L^3}} \, {\mathbf \epsilon}_{{\bf k} \lambda} \, a_{{\bf k} \lambda} \, {\rm e}^{{\rm i} {\bf k} \cdot {\bf R}} + {\rm H.c.} \, , \notag \\
\textbf{E}_{\rm L} (\textbf{R},t)&=& \textbf{E}_0 \, {\rm e}^{{\rm i}(\textbf{k}_{\rm L} \cdot\textbf{R}-\omega_{\rm L} t)} + {\rm c.c.}  
\end{eqnarray}
with $L^3$ being the quantisation volume of the free radiation field and $ {\mathbf \epsilon}_{{\bf k} \lambda}$ being a unit length polarisation vector orthogonal to ${\bf k}$. Moreover, $\textbf{E}_0$, $\textbf{k}_{\rm L}$, and $\omega_{\rm L}$ are the amplitude, the wave vector of length $k_{\rm L}$, and the frequency of the applied laser field, respectively. 

\subsection{The displacement operator} \label{symi}

In the following, we assume that the incoming laser field has the same direction as the quantised motion of the trapped particle, since this maximises the effect of the cooling laser. Considering this motion as quantised with the phonon annihilation operator $b$ from above, hence implies
\begin{eqnarray} \label{2.21}
\textbf{k}_{\rm L} \cdot \textbf{R} &=& \eta \left(b + b^\dagger \right) \, , 
\end{eqnarray}
where the Lamb-Dicke parameter $\eta $ is a measure for the steepness of the effective trapping potential seen by the ion \cite{Stenholm2}. Notice that Eq.~(\ref{2.21}) applies as long as the trapping potential seen by the atom does not depend on its respective electronic state. This means, the following calculations apply to a trapped ion and to a trapped neutral atom with a magical wavelength \cite{magic}. 

Taking Eq.~(\ref{2.21}) into account, we find that the laser Hamiltonian is a function of the particle displacement operator \cite{Knight}
\begin{eqnarray}\label{2.22}
D ({\rm i}\eta) &\equiv & {\rm e}^{- {\rm i} \eta (b+b^\dagger)} 
\end{eqnarray}
which is a unitary operator with
\begin{eqnarray}\label{2.222}
D ({\rm i}\eta) \, b \, D({\rm i}\eta)^\dagger &=& b + {\rm i} \eta \, , \nonumber \\
D ({\rm i}\eta)^\dagger \, b \, D ({\rm i}\eta) &=& b - {\rm i} \eta \, .
\end{eqnarray}
Using the polar coordinates $\vartheta$ and $\varphi$, putting the $z$-axis in the direction of the cooling laser, and writing the general wave vector ${\bf k}$ of length $k$ as 
\begin{eqnarray}\label{2.4}
{\bf k} &=& k \left( \begin{array}{c} \sin \vartheta \cos \varphi \\ \sin \vartheta \sin \varphi \\ \cos \vartheta \end{array} \right) 
\end{eqnarray}
we find that
\begin{eqnarray} \label{2.21x}
\textbf{k} \cdot \textbf{R} &=& k \sin \vartheta \left[ R_x \cos \varphi + R_y \sin \varphi \right] \notag \\
&& + {\eta k \cos \vartheta \over k_{\rm L}} \left(b + b^\dagger \right) \, , 
\end{eqnarray}
where $R_x$ and $R_y$ are the $x$ and the $y$ component of the vector ${\bf R}$. Writing the Hamiltonian $H_{\rm dip}$ in Eq.~(\ref{HL1}) as a function of displacement operators, it becomes
\begin{eqnarray}\label{2.4}
H_{\rm dip} (t) &=& e \left[ \textbf{D}_{01} \, \sigma^{-} + {\rm H.c.} \right] \cdot  \Bigg[ \textbf{E}_0^* \, D({\rm i}\eta) \, {\rm e}^{ {\rm i} \omega_{\rm L} t}  \notag \\
&& - {\rm i} \sum_{{\bf k} \lambda} \sqrt{{\hbar \omega_k \over 2 \epsilon_0 L^3}} \, {\mathbf \epsilon}_{{\bf k} \lambda} \, a_{{\bf k} \lambda}^\dagger \, D \left({{\rm i} \eta k \cos \vartheta \over k_{\rm L}} \right) \notag \\
&& \times {\rm e}^{ - {\rm i} k \sin \vartheta  \left[ R_x \cos \varphi + R_y \sin \varphi \right]}  \Bigg] + {\rm H.c.}
\end{eqnarray}
This equation shows that the cooling laser establishes a coupling between the electronic states $|0 \rangle$ and $|1 \rangle$ of the trapped particle and its quantised motion. As we shall see below, the coupling to the free radiation field is the origin of spontaneous emission and recoil effects.

\subsection{Interaction picture} \label{IP}

Before continuing our derivation of the master equation, it is convenient to transform the Hamiltonian $H$ in Eq.~(\ref{1.23}) into an interaction picture. To do so, we choose
\begin{eqnarray} \label{intpic}
H_0 &=& \hbar\omega_{\rm L} \, \sigma^{+}\sigma^{-} + H_{\rm field} 
\end{eqnarray}
with $H_{\rm field}$ as in Eq.~(\ref{1.23b}). Neglecting relatively fast oscillating terms as part of the usual rotating wave approximation, the interaction Hamiltonian $H_{\rm I}$,
\begin{eqnarray} \label{trafo}
H_{\rm I}=U^{\dagger}_0(t,0) \, (H-H_0) \, U_0(t,0) \, ,
\end{eqnarray}
becomes 
\begin{eqnarray}\label{2.8}
H_{\rm I} &=& \sum_{{\bf k} \lambda} \hbar g_{{\bf k} \lambda} \, \sigma^{-} a_{{\bf k} \lambda}^\dagger \, D \left({{\rm i} \eta k \cos \vartheta \over k_{\rm L}} \right) \notag \\
&& \times {\rm e}^{ - {\rm i} k \sin \vartheta  \left[ R_x \cos \varphi + R_y \sin \varphi \right]} \, {\rm e}^{{\rm i} (\omega_k - \omega_{\rm L}) t} \notag \\
&& + {1 \over 2} \hbar \Omega \, D({\rm i}\eta) \sigma^- +{\rm H.c.}+\hbar\Delta \sigma^+\sigma^- +\hbar\nu \, b^{\dagger}b \, . ~~~~~
\end{eqnarray}
Here $\Delta$ denotes the detuning between the laser and the relevant atomic transition and $\Omega$ and $g_{{\bf k} \lambda}$, 
\begin{eqnarray} \label{symi2}
\Omega &=& {2 e \, \textbf{D}_{01} \cdot {\bf E}_0^* \over \hbar} \, , \notag \\
g_{{\bf k} \lambda} &=& - {\rm i} e \, \sqrt{{\omega_k \over 2 \hbar \epsilon_0 L^3}} \, \textbf{D}_{01} \cdot {\mathbf \epsilon}_{{\bf k} \lambda}
\end{eqnarray}
are the usual laser Rabi frequency and the atom-field coupling constant.

\subsection{Spontaneous emission and recoil}

The interaction Hamiltonian $H_{\rm I}$ in Eq.~(\ref{2.8}) is now the starting point for the usual derivation of the master equations. Suppose the state of the laser-driven trapped particle is at $t=0$ given by the density matrix $\rho$, while the free radiation field is in its vacuum state $|0 \rangle$. Taking this into account, the density matrix $\rho (\Delta t)$ of the particle at time $\Delta t$ can be written as \cite{Hegerfeldt}
\begin{eqnarray} \label{ME1}
\rho (\Delta t) &=& U_{\rm cond} (\Delta t,0) \, \rho \, U_{\rm cond}^\dagger (\Delta t,0) + {\cal R}(\rho) \Delta t \, , ~~
\end{eqnarray}
where $\Delta t$ denotes the typical response time of the environment, i.e.~the typical time it takes the environment to absorb a photon from the free radiation field, and where 
\begin{eqnarray} \label{long}
U_{\rm cond} (\Delta t,0) &=& \langle 0| \, U_{\rm I} (\Delta t,0) \, |0 \rangle \, , \notag \\
{\cal R}(\rho) &=& \lim_{\Delta t \to 0} {1 \over \Delta t} \sum_{{\bf k} \lambda} \langle 1_{{\bf k} \lambda}| \, U_{\rm I} (\Delta t,0) \, \rho \notag \\
&& \hspace*{0.7cm} \otimes |0 \rangle \langle 0| \, U_{\rm I}^\dagger (\Delta t,0) \, |1_{{\bf k} \lambda} \rangle \, . 
\end{eqnarray}
The first term in Eq.~(\ref{ME1}) describes the subensemble with no photon emission in $\Delta t$. The second term in this equation is the unnormalised state of the subensemble with an emission in $\Delta t$ \cite{Hegerfeldt}. Taking the time derivation of $\rho (\Delta t)$ on the coarse grained time scale $\Delta t$ into account, we obtain the usual master equation
\begin{eqnarray} \label{master}
\dot \rho &=& - {{\rm i} \over \hbar} \, \left[ H_{\rm cond} \, \rho - \rho \, H_{\rm cond}^\dagger \right] + {\cal R}(\rho) \, , 
\end{eqnarray}
in Lindblad form, where $H_{\rm cond}$ is a non-Hermitian conditional Hamiltonian with $U_{\rm cond}$ being the corresponding no-photon time evolution operator.

In order to calculate $H_{\rm cond}$ and ${\cal R}(\rho)$, one usually exploits second order perturbation theory. Since the displacement operator $D({\rm i} \eta)$ is a unitary operator, i.e.
\begin{eqnarray}
D({\rm i} \eta) D({\rm i} \eta)^\dagger = D({\rm i} \eta)^\dagger D({\rm i} \eta) &=&1 \, , 
\end{eqnarray}
the derivation of the conditional Hamiltonian $H_{\rm cond}$ remains exactly the same as in the case, where the motion of the particle is not quantised. This means, we find that
\begin{eqnarray}\label{2.8cond}
H_{\rm cond} &=& {1 \over 2} \hbar \Omega \, D({\rm i}\eta) \sigma^- +{\rm H.c.}+\hbar\Delta \, \sigma^+\sigma^- \notag \\
&& + \hbar\nu \, b^{\dagger}b - {{\rm i} \over 2} \hbar \Gamma \, \sigma^+ \sigma^- \, , ~~~~~
\end{eqnarray}
where the spontaneous decay rate $\Gamma$ of the excited electronic state $|1 \rangle$ is given by
\begin{eqnarray} \label{symi3}
\Gamma &=& {e^2 \omega_0^3 \over 3 \pi \epsilon_0 \hbar c^3} \,  |{\bf D}_{01}|^2 \, .
\end{eqnarray}
However, the reset operator ${\cal R}(\rho)$ now contains recoil terms. Proceeding as described in App.~\ref{app}, we find that 
\begin{eqnarray} \label{longfinal}
{\cal R}(\rho) &=& {3 \Gamma \over 8} \, \int_{-1}^1 {\rm d} \zeta \, \sigma^- D({\rm i} \eta \zeta ) \, \rho \, D({\rm i} \eta \zeta )^\dagger \sigma^+ \notag \\
&& \times \left[ \, 1 + |d_3|^2 + \left( 1 - 3 |d_3|^2 \right) \zeta^2 \, \right] \, ,
\end{eqnarray}
where $d_3$ denotes the $z$-component of the normalised dipole vector ${\bf D}_{01}/|{\bf D}_{01}|$. The above reset operator is different from the one often used in the literature \cite{Stenholm2,sideband2,Stenholm4}. The reason for this is that the authors of these references only consider the case where $d_3 = 0$, as it applies to certain atomic level schemes and laser configurations. The above reset operator ${\cal R}(\rho)$ is more general. It is also consistent with the reset operator of a free particle whose motion is not quantised. In this case, the displacement operator $D({\rm i} \eta \cos \vartheta )$ becomes a number and the integration over $\xi$ results indeed in ${\cal R}(\rho) = \Gamma \, \sigma^- \, \rho \sigma^+$. This expression is independent of $d_3$, as it should.

\section{Cooling equations} \label{standard}

The master equations which we derived in the previous section can now be used to derive differential equations for expectation values, so-called rate or cooling equations. However, this is not a straightforward task due to the presence of the displacement operator $D$ in Eq.~(\ref{2.8cond}). To overcome this problem, we now introduce two new operators $x$ and $y$ which replace the particle and the phonon operators $\sigma^-$ and $b$, respectively. Both operators describe neither electronic excitations nor phonons but provide nevertheless a natural description of trapped particles. 

\subsection{Transformation of the Hamiltonian}

To simplify the Hamiltonian $H_{\rm I}$ in Eq.~(\ref{2.4}), we now introduce a new annihilation operator $x$ as 
\begin{eqnarray}\label{25}
x &\equiv& D({\rm i} \eta) \, \sigma^- \, . 
\end{eqnarray} 
Using the commutator relation (\ref{comms}) and the unitarity of $D$ in Eq.~(\ref{2.22}), i.e.~the fact that
\begin{eqnarray}\label{2.22}
D ({\rm i}\eta) D ({\rm i}\eta)^\dagger =  D ({\rm i}\eta) D ({\rm i}\eta)^\dagger &=& 1 \, ,
\end{eqnarray}
one can show that $x$ obeys the commutator relation
\begin{eqnarray}\label{25}
\left[ x, x^\dagger \right] &=& 1 - 2 \, x^\dagger x \, . 
\end{eqnarray} 
The operator $x$ differs from $\sigma^-$ by the fact that its application not only transforms $|1 \rangle$ into $|0 \rangle$ but also induces a kick which displaces the particle.

Eqs.~(\ref{comms}), (\ref{2.222}), and (\ref{25}) can now be used to derive for example the commutator relations
\begin{eqnarray}\label{28}
\left[ x, b \right] = - \left[ x, b^\dagger \right] &=& {\rm i} \eta \, x \, , \nonumber \\
\left[ x^{\dagger}, b \right] = - \left[ x^\dagger, b^\dagger \right] &=& - {\rm i} \eta \, x^\dagger \, .
\end{eqnarray} 
These can then be used to show that
\begin{eqnarray}\label{29}
&& \left[ x, b^\dagger b \right] = - {\rm i} \eta \, x (b - b^\dagger) - \eta^2 \, x \, , \nonumber \\
&& \left[ x^\dagger, b^\dagger b \right] = {\rm i} \eta (b - b^\dagger) x^\dagger + \eta^2 \, x^\dagger \, , \nonumber \\
&& \left[ x^\dagger x, b \right] \, = \, \left[ x^\dagger x, b^\dagger \right]  \, = \, \left[ x^\dagger x, b^\dagger b \right] = 0 \, .  
\end{eqnarray} 
The operators $x$ and $b$ and functions of them do not commute in general. This means, although substituting Eq.~(\ref{25}) into the interaction Hamiltonian $H_{\rm I}$  into Eq.~(\ref{2.4}), simplifies it a lot, we do not yet have a Hamiltonian which can be analysed easily.

To overcome this problem we now introduce another operator $y$ as 
\begin{eqnarray} \label{25y}
y &\equiv & b - {\rm i} \eta \, x^\dagger x \, .
\end{eqnarray}
Using the commutator relations in Eq.~(\ref{29}), one can show that $y$ is a bosonic operator which obeys the commutator relation
\begin{eqnarray}\label{26c}
\left[ y, y^\dagger \right] &=& 1 \, .
\end{eqnarray} 
Moreover, we now have 
\begin{eqnarray}\label{33}
\left[ x, y \right] = \left[ x^\dagger, y \right] &=& 0 
\end{eqnarray} 
which can be checked using the commutator relations in Eqs.~(\ref{28}) and (\ref{26c}).

Using the notation introduced in this section, the conditional Hamiltonian $H_{\rm cond}$ in Eq.~(\ref{2.8cond}) 
and the reset operator ${\cal R}(\rho)$ in Eq.~(\ref{longfinal}) become
\begin{eqnarray} \label{35}
H_{\rm cond} &=& {1\over2}\hbar \Omega \, \left( x + x^\dagger \right) - {\rm i} \hbar \eta \nu \, x^\dagger x (y -y^\dagger)  \nonumber \\
&& + \hbar \left( \Delta + \eta^2 \nu \right) \, x^{\dagger} x + \hbar \nu \, y^{\dagger} y - {{\rm i} \over 2} \hbar \Gamma \, x^\dagger x \, , \nonumber \\
{\cal R}(\rho) &=& {3 \Gamma \over 8} \, \int_{-1}^1 {\rm d} \zeta \, x D({\rm i} \eta (1 - \zeta) )^\dagger \, \rho \, D({\rm i} \eta ( 1 - \zeta) ) x^\dagger \notag \\
&& \times \left[ \, 1 + |d_3|^2 + \left( 1 - 3 |d_3|^2 \right) \zeta^2 \, \right] \, .
\end{eqnarray}
The new conditional Hamiltonian $H_{\rm cond}$ no longer contains any exponential terms. 

\subsection{Time evolution of expectation values} \label{REV}

The time derivative of the expectation value of a time-independent operator $A$ equals 
\begin{eqnarray}
\langle \dot A \rangle &=& \mbox{Tr} (A \dot{\rho}) \, .
\end{eqnarray}
The master equation in Eq.~(\ref{master}) hence implies that
\begin{eqnarray} \label{dotA}
\langle \dot A \rangle &=& -{{\rm i} \over \hbar} \, \langle A H_{\rm cond} - H_{\rm cond}^\dagger A \rangle \nonumber \\
&& +  {3 \Gamma \over 8} \, \int_{-1}^1 {\rm d} \zeta \, \langle x^\dagger D({\rm i} \eta (1 - \zeta)) \, A \, D({\rm i} \eta (1 - \zeta))^\dagger x \rangle \notag \\
&& \times \left[ \, 1 + |d_3|^2 + \left( 1 - 3 |d_3|^2 \right) \zeta^2 \, \right] 
\end{eqnarray}
with $H_{\rm cond}$ as in Eq.~(\ref{35}). This equation describes the time evolution of the expectation value $\langle A \rangle$ within the interaction picture which we introduced previously in Section \ref{IP}. 

In the following we are especially interested in the time evolution of the mean phonon number $m$,
\begin{eqnarray} \label{ns}
m &\equiv & \langle b^{\dagger} b \rangle \, .
\end{eqnarray} 
Using Eqs.~(\ref{25}) and (\ref{25y}), we find that this expression is the same as
\begin{eqnarray} \label{ns3}
m &\equiv & n_2 - \eta \, k_{12} + \eta^2 \, n_1 
\end{eqnarray} 
with $n_1$, $n_2$, and $k_{12}$ defined as
\begin{eqnarray} \label{coherences}
n_1 \equiv \langle x^{\dagger} x \rangle \, , ~~ n_2 \equiv \langle y^{\dagger} y \rangle \, , ~~
k_{12} \equiv {\rm i}  \, \langle x^\dagger x (y - y^\dagger ) \rangle \, . ~
\end{eqnarray}
To predict the time evolution of $m$, we need to evaluate the time evolution of these three expectation values. As we shall see below, we only obtain a closed set of cooling equations, if we consider in addition the expectation values
\begin{eqnarray} \label{coherences}
&& \hspace*{-0.3cm} k_7 \equiv \langle y + y^\dagger \rangle \, , ~~
k_8 \equiv {\rm i} \, \langle y - y^\dagger \rangle \, , ~~ k_9 \equiv \langle y^2 + y^{\dagger \, 2} \rangle \, , \nonumber \\ 
&& \hspace*{-0.3cm} k_{10} \equiv {\rm i} \, \langle y^2 - y^{\dagger \, 2} \rangle \, , ~~ 
k_{11} \equiv \langle x^\dagger x (y + y^\dagger ) \rangle 
\end{eqnarray}
and the expectation values defined in App.~\ref{appA}. Since all of these variables are expectation values of Hermitian operators, they are real and  their time evolution is given by real differential equations. 

Let us first have a look at the $y$ operator expectation values $n_2$ and $k_7$ to $k_{10}$. Using the master equation in Eq.~(\ref{dotA}) to calculate the time derivatives of these five variables, we find that 
\begin{eqnarray} \label{4888}
\dot n_2 &=& \eta \nu \, k_{11} - \eta \Gamma \, k_{12} + \eta^2 \theta \Gamma \, n_1 \, , \notag \\
\dot k_7 &=& 2 \eta \nu \, n_1 - \nu \, k_8 \, , \nonumber \\ 
\dot k_8 &=& \nu \, k_7 - 2 \eta \Gamma \, n_1 \, , \nonumber \\
\dot k_9 &=& - 2 \nu \, k_{10} + 2 \eta \nu \, k_{11} + 2 \eta \Gamma \, k_{12} - 2 \eta^2 \theta \Gamma \, n_1 \, , ~~~~ \nonumber \\
\dot k_{10} &=& 2 \nu \, k_9 + 2 \eta \nu \, k_{12} - 2 \eta \Gamma \, k_{11} \, .
\end{eqnarray}
The factor $\theta$ in this equation,
\begin{eqnarray}
\theta \equiv {1 \over 5} (7 - |d_3|^2) \, ,
\end{eqnarray}
depends explicitly on the direction of the emitting dipole moment. It relates to the parameter $\alpha$ used in previous papers \cite{Stenholm2,sideband2,Stenholm4,ions} via the equation
\begin{eqnarray} \label{alpha}
\theta &=& 1 + \alpha - {1 \over 5} \, |d_3|^2 \, .
\end{eqnarray}
The time derivatives of the relevant $x$ and mixed operator expectation values can be found in App.~\ref{appB}. Below we simplify these equations with the help of an adiabatic elimination of all relatively fast evolving variables.

\subsection{Weak confinement regime} \label{effeq}

We now have a closer look at the case, where the trapped particle experiences a relatively weak trapping potential and where the Lamb-Dicke parameter $\eta$ is much smaller than one. More concretely we assume in the following that
\begin{eqnarray} \label{weak}
\nu \ll \Gamma ~~ {\rm and} ~~ \eta \ll 1\, .
\end{eqnarray}
In addition we assume that the Rabi frequency $\Omega$ and the detuning $\Delta$ are at most comparable to $\Gamma$ and definitely not much larger. However, notice that we do {\em not} demand that $\Omega$ is much smaller than $\Gamma$. A closer look at the cooling equations in App.~\ref{appB} shows that this choice of parameters causes the $y$ operator expectation values $n_2$ and $k_7$ to $k_{10}$ to evolve on a much slower time scale than all other relevant expectation values. The reason for this is that these variables are all $x$ or mixed operator expectation values which decay with the spontaneous atomic decay rate $\Gamma$. 

Taking this into account and eliminating $n_1$, $k_1$, $k_2$, and $k_{13}$ to $k_{24}$ adiabatically from the system dynamics, we obtain a closed set of five effective cooling equations which applies after a relatively short transition time and which can be written as 
\begin{eqnarray} \label{eff}
\big( \dot n_2 , \dot k_7 , \dot k_8 , \dot k_9 , \dot k_{10} \big)^{\rm T} 
&=& M \left( n_2 , k_7 , k_8 , k_9 , k_{10} \right)^{\rm T}  \nonumber \\
&& + \left( \beta_1 , \beta_2 , \beta_3 , \beta_4 , \beta_5 \right)^{\rm T} \, . ~~~
\end{eqnarray}
A closer look at Eq.~(\ref{4888}) shows that the time derivatives of the $y$ operator expectation values $n_2$, and $k_7$ to $k_{10}$ depend only on $n_1$, $k_{11}$, and $k_{12}$. The calculation of the $5 \times 5$ matrix  $M$ therefore only requires the calculation of $n_1$, $k_{11}$, and $k_{12}$ which can be found in App.~\ref{appB}. Substituting Eqs.~(\ref{zeroth}), (\ref{k11120}), (\ref{zeroth2}), and (\ref{k11121}) into Eq.~(\ref{4888}), we find that $M$ can be written as
\begin{eqnarray} \label{eff33}
M &=& \left( \begin{array}{ccccc} \alpha_{11}^{(2)} & \alpha_{12}^{(1)} & \alpha_{13}^{(1)} & \alpha_{14}^{(2)} & 0 \\ 
0 & 0 & - \nu & 0 & 0 \\
0 & \nu & \alpha_{33}^{(2)} & 0 & 0 \\
\alpha_{41}^{(2)} & \alpha_{42}^{(1)} & \alpha_{43}^{(1)} &\alpha_{44}^{(2)} & - 2 \nu \\
0 & \alpha_{52}^{(1)} & \alpha_{53}^{(1)} & 2 \nu & \alpha_{55}^{(2)} \end{array} \right) . ~~~
\end{eqnarray}
The first order matrix elements $\alpha_{ij}^{(1)}$ in this equations are given by
\begin{eqnarray} \label{eff4}
&& \alpha_{12}^{(1)} = - {2 \eta \nu \Omega^2 \over \mu^4} \, (\Gamma^2 - 4 \Delta^2 - \Omega^2) \, , ~~
\alpha_{13}^{(1)} = - {\eta \Gamma \Omega^2 \over \mu^2} \, , ~~~~ \nonumber \\  
&& \alpha_{42}^{(1)} = {4 \eta \nu \Omega^2 \over \mu^4} \, (\Omega^2 + 2 \Gamma^2) \, , ~~  
\alpha_{43}^{(1)} = {2 \eta \Gamma \Omega^2 \over \mu^2} \, , \nonumber \\
&& \alpha_{52}^{(1)} = - \alpha_{43}^{(1)} \, , ~~ \alpha_{53}^{(1)} = \alpha_{42}^{(1)} 
\end{eqnarray}
with $\mu^2$ defined as in Eq.~(\ref{eff44}). The non-zero matrix elements $\alpha_{ij}^{(2)}$ of $M$ in second order in $\eta$ and in first order in $\nu$ are given by  
\begin{eqnarray} \label{eff88xxx}
&& \alpha_{11}^{(2)} = \alpha_{33}^{(2)} = \alpha_{44}^{(2)} = \alpha_{55}^{(2)} = - {16 \eta^2 \nu \Delta \Gamma \Omega^2 \over \mu^4} \, , ~~ \nonumber \\
&& \alpha_{14}^{(2)} = {8 \eta^2 \nu \Delta \Gamma \Omega^2 \over \mu^4} \, , ~~
\alpha_{41}^{(2)} = {32 \eta^2 \nu \Delta \Gamma \Omega^2 \over \mu^4} \, . ~~ 
\end{eqnarray}
Proceeding as above, one can show that $\beta_1 = \beta_1^{(2)}$ is up to second order in $\eta$ to a very good approximation given by  
\begin{eqnarray} \label{beta1}
\beta_1^{(2)} &=& {\eta^2 \Gamma \Omega^2 \over \mu^2} \, \theta \, . ~~
\end{eqnarray}
Moreover, we find that the coefficients $\beta_2$ to $\beta_5$ in Eq.~(\ref{eff}) equal  
\begin{eqnarray} \label{betas}
&& \beta_2^{(1)} = {2 \eta \nu \Omega^2 \over \mu^2} \, , ~~
\beta_3^{(1)} = - {2 \eta \Gamma \Omega^2 \over \mu^2} \, , \notag \\
&&  \beta_4^{(1)} = \beta_5^{(1)} = 0 
\end{eqnarray}
in first order in $\eta$. We now have a closed set of five differential equations which can be used to analyse the time evolution of the $y$ operator expectation values analytically and numerically. Notice that the weak confinement regime which we introduced in Eq.~(\ref{weak}) does not allow for the adiabatic elimination of the $y$ operator coherences $k_7$ to $k_{10}$, since these evolve in general on the same time scale as the $y$ particle population $n_2$.

\subsection{Strong confinement regime} \label{effeq2}

Let us now have a closer look at the parameter regime where the phonon frequency $\nu$ and the detuning $\Delta$ exceed the spontaneous decay rate $\Gamma$ and the Rabi frequency $\Omega$ by at least one order of magnitude, 
\begin{eqnarray} \label{cond}
\Omega, \, \Gamma &\ll & \nu, \, \Delta \, , ~~ {\rm while} ~~ \eta \ll 1\, .
\end{eqnarray}
In this so-called strong confinement regime, the time scale separation in the dynamics of the trapped particle is different than in the previous subsection. The adiabatic elimination performed in the previous section does not apply. However, at least at the end of the cooling process when $n_2$ is already very small, we can assume that the expectation values $n_1$, $n_4$, $k_1$, $k_2$, and $k_7$ to $k_{24}$ evolve much faster than the $y$ operator population $n_2$. This means, we can simplify the system dynamics via an adiabatic elimination of all expectation values other than $n_2$. Doing so (cf.~App.~\ref{appF}), we obtain the effective cooling equation 
\begin{eqnarray} \label{strong2}
\dot n_2 &=& - \gamma_{\rm c} \, n_2 + c \, .
\end{eqnarray}
The frequencies $\gamma_{\rm c}$ and $c$ in this equation are given by
\begin{eqnarray} \label{eff88strong}
\gamma_{\rm c} &=& {\eta^2 \Gamma \Omega^2 \over 4 (\Delta - \nu)^2} - {\eta^2 \Gamma \Omega^2 \over 4 (\Delta + \nu)^2} \, , \notag \\
c &=&{\eta^2 \Gamma \Omega^2 \over 4 \Delta^2}  \left[ \theta + {\Delta^2 \over (\Delta + \nu)^2} -1 \right]
\end{eqnarray}
up to second order in $\eta$. It is obvious that $\gamma_{\rm c}$ is the effective cooling rate for strongly confined particles. 

\section{Stability analysis} \label{stability}

\noindent \begin{figure*}[t]
\begin{minipage}{2\columnwidth}
\begin{center}
\resizebox{\columnwidth}{!}{\rotatebox{0}{\includegraphics{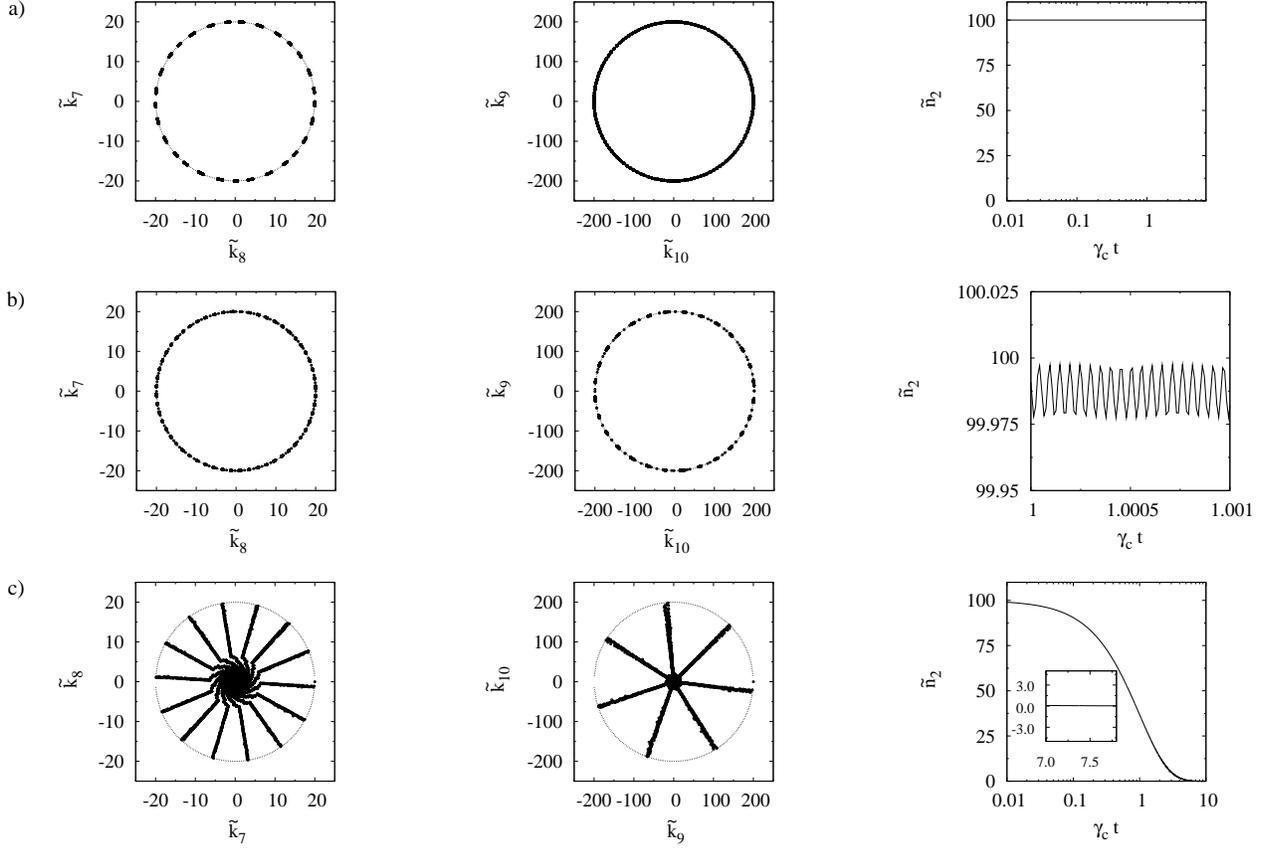}}} 
\end{center}
\caption{Diagrams illustrating the time evolution of the expectation values $\tilde k_7$ to $\tilde k_{10}$, and $\tilde n_2$ for $\Delta = 0.5 \, \Gamma$, $\nu = 0.1 \, \Gamma$, $\Omega = 0.01 \, \Gamma$, and $d_3 = 0$. It is assumed that the $y$ particles are initially in a coherent state. In (a), only terms in zeroth order in $\eta$ have been taken into account. As expected, we find that the mean phonon number remains constant in time. The coherences $\tilde k_7$ to $\tilde k_{10}$ evolve such that their points in the respective phase diagrams lie on circles. This means that the $y$ particles remain in a coherent state. In (b), only terms in first order in $\eta$ have been taken into account. All five eigenvalues of $M$ are still either zero or purely imaginary which is why there is still no reduction of $\tilde n_2$. In (c), also the second order terms in $\eta$ are taken into account. The coherences $\tilde k_7$ to $\tilde k_{10}$ now evolve towards zero. We now observe an exponential decrease of $\tilde n_2$. This implies a reduction of the mean phonon number $m$ in zeroth order in $\eta$, ie.~cooling.} \label{big}
\end{minipage}
\end{figure*}

In the strong confinement regime, the cooling process is governed by a single differential equation (cf.~Eq.~(\ref{strong2})) with an always positive cooling rate $\gamma_{\rm c}$ and an always positive stationary state phonon number $m_{\rm ss}$. This means, in the strong confinement regime, the trapped particle always reaches its stationary state. However, it is not clear whether the same applies in the weak confinement regime, where the cooling process is described by five linear differential equations (cf.~Eq.~(\ref{eff})). In this section, we therefore have a closer look at the dynamics induced by these equations. 

To do so, we introduce the shifted $y$ operator expectation values 
\begin{eqnarray} \label{effxx}
\big( \tilde n_2 , \tilde k_7 , \tilde k_8 , \tilde k_9 , \tilde k_{10} \big)^{\rm T} 
&\equiv & \left( n_2 , k_7 , k_8 , k_9 , k_{10} \right)^{\rm T} \nonumber \\
&& + M^{-1} \left( \beta_1 , \beta_2 , \beta_3 , \beta_4 , \beta_5 \right)^{\rm T} . ~~~~~~
\end{eqnarray}
This definition means that the tilde and the non-tilde variables differ only by the stationary state solutions of the non-tilde expectation values. Substituting Eq.~(\ref{effxx}) into Eq.~(\ref{eff}), we find that
\begin{eqnarray} \label{effyy}
\big( \dot{\tilde{n}}_2 , \dot{\tilde k}_7 , \dot{\tilde k}_8 , \dot{\tilde k}_9 , \dot{\tilde k}_{10} \big)^{\rm T} 
= M \left( \tilde n_2 , \tilde k_7 , \tilde k_8 , \tilde k_9 , \tilde k_{10} \right)^{\rm T} \, . ~~
\end{eqnarray}
The stationary state solution of these effective cooling equations is the trivial one with all variables equal to zero. However, notice that the corresponding stationary state is only reached, if all eigenvalues of $M$ have negative eigenvalues. 

\subsection{Time evolution for $\eta = 0$}

We first have a closer look at the time evolution of the $y$ operator expectation values for $\eta=0$. One can easily check that the eigenvalues of the matrix $M$ in Eq.~(\ref{eff33}) are in this case simply given by 
\begin{eqnarray} \label{eff55}
\lambda_1 = 0 \, , ~~ \lambda_{2,3} = \mp {\rm i} \nu \, , ~~ \lambda_{4,5} = \mp 2 {\rm i} \nu \, .  
\end{eqnarray}
Taking this into account and solving Eq.~(\ref{effyy}) analytically, we find that
\begin{eqnarray} \label{eff552}
\tilde n_2 (t) &=& \tilde n_2(0) \, , \nonumber \\  
\left( \begin{array}{c} \tilde k_7 (t) \\ \tilde k_8 (t) \end{array} \right) &=& \left( \begin{array}{rr} \cos \nu t & - \sin \nu t \\ \sin \nu t & \cos \nu t \end{array} \right) \left( \begin{array}{c} \tilde k_7 (0) \\ \tilde k_8 (0) \end{array} \right) \, , \nonumber \\
\left( \begin{array}{c} \tilde k_9 (t) \\ \tilde k_{10} (t) \end{array} \right) &=& \left( \begin{array}{rr} \cos 2 \nu t & - \sin 2 \nu t \\ \sin 2 \nu t & \cos 2 \nu t \end{array} \right) \left( \begin{array}{c} \tilde k_9 (0) \\ \tilde k_{10} (0) \end{array} \right) \, . ~~~~
\end{eqnarray}
This behaviour is illustrated in Fig.~\ref{big}(a) which shows numerical solutions of the effective cooling equations in Eq.~(\ref{effyy}) for the case where the $y$ particles are initially in a coherent state. The first two phase diagrams show $\tilde k_8$ and $\tilde k_{10}$ as functions of $\tilde k_7$ and $\tilde k_9$, respectively. The fact that all points lie on a circle means that the state of the $y$ particles remains (at least approximately) coherent throughout the cooling process. Fig.~\ref{big}(a) moreover shows that the $y$ particle population $\tilde n_2$ and therefore also the mean phonon number $m$ (cf.~Eq.~(\ref{ns3})) in zeroth order in $\eta$ remains constant in time. This is exactly as one would expect. There cannot be any cooling without a coupling between the electronic and the vibrational states of the trapped particle. Higher order corrections in $\eta$ have to be taken into account. 

\subsection{First order corrections}

Calculating the eigenvalues of $M$ in Eq.~(\ref{eff33}) up to first order in $\eta$, we obtain again Eq.~(\ref{eff55}). All of them are either zero or imaginary. However, there are some small corrections to the eigenvectors in first order in $\eta$. As a result, the shifted $y$ particle population $\tilde n_2$ remains no longer constant in time but oscillates on the time scale given by the phonon frequency $\nu$. This is illustrated in Fig.~\ref{big}(b) which shows a numerical solution of the effective cooling equations in Eq.~(\ref{effyy}) with all first order corrections in $\eta$ taken into account. However, since the eigenvalues of $M$ have no real parts, $\tilde n_2$ does not reach its stationary state value. No cooling occurs.

\subsection{Second order corrections}

Calculating again the eigenvalues of the matrix $M$ in Eq.~(\ref{eff33}) but now taking also the terms in second order in $\eta$ into account, we find that 
\begin{eqnarray} 
\lambda_1 &=& \alpha_{11}^{(2)} \, , \notag \\
\lambda_{2,3} &=& {1\over 2} \, \alpha_{11}^{(2)} \mp {{\rm i} \over 2} \sqrt { 4 \nu^2 - \left(\alpha_{11}^{(2)} \right)^2 } \, , \notag \\
\lambda_{4,5} &=& \alpha_{11}^{(2)} \mp {\rm i} \sqrt{ 4 \nu^2 - \alpha_{14}^{(2)} \alpha_{41}^{(2)}} \, .
\end{eqnarray}
Since the matrix element $\alpha_{11}^{(2)}$ is always negative, all five eigenvalues of $M$ have negative real parts. This means, all tilde variables are damped away and tend to zero on the time scale given by $1/\alpha_{11}^{(2)}$. This is illustrated in Fig.~\ref{big}(c) which illustrates a numerical solution of Eq.~(\ref{effyy}).

Since the $y$ coherences $k_7$ to $k_{10}$ do not increase in time but oscillate instead with a slowly decreasing amplitude around constant values, the cooling process remains stable and the trapped particle eventually reaches its stationary state. This observation is taken into account in the following section, where we analyse the cooling process in more detail. It allows us to replace the coherences $k_7$ to $k_{10}$ by their time averages. This means, in the weak confinement regime, the calculations in the following section apply only towards the end of the cooling process, i.e.~after a transition time of the order of$1/\alpha_{11}^{(2)}$.

\section{Cooling rates and phonon numbers} \label{process}

\noindent \begin{figure*}[t]
\begin{minipage}{2\columnwidth}
\begin{center}
\resizebox{\columnwidth}{!}{\rotatebox{0}{\includegraphics{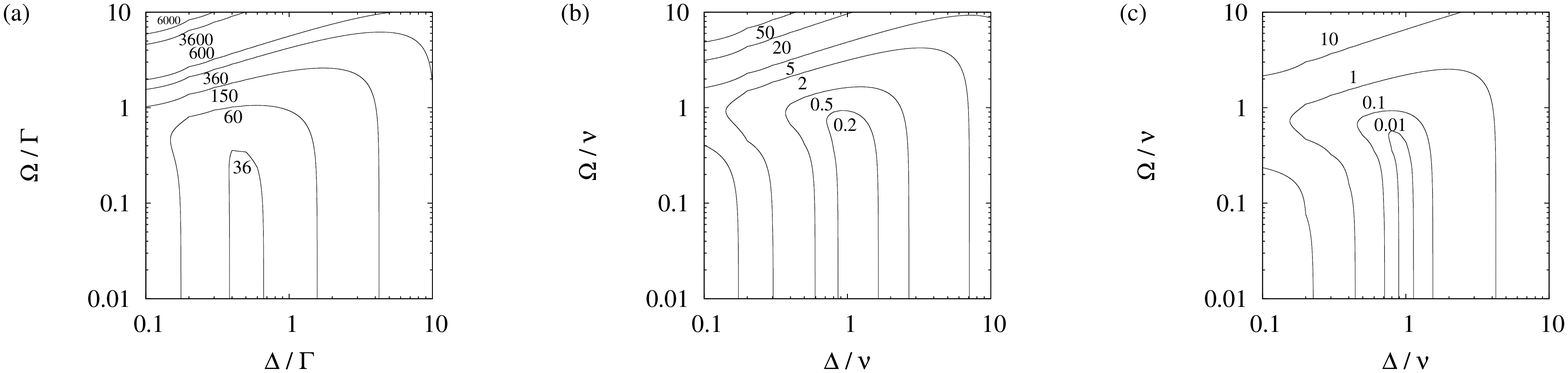}}} 
\end{center}
\caption{Logarithmic contour plot of the stationary state phonon number $m_{\rm ss}$ in Eq.~(\ref{mss}) as a function of the laser parameters $\Omega$ and $\Delta$ for (a) $\nu = 0.01 \, \Gamma$, (b) $\Gamma = \nu$, and (c) $\Gamma = 0.01 \, \nu$.} \label{contour}
\end{minipage}
\end{figure*}

Taking the results of the previous section into account, replacing the $y$ operator coherences $k_7$ to $k_{10}$ by their time averages, and adiabatically eliminating all rapidly evolving expectation values, we obtain an effective cooling equation of the same form as the effective cooling equation in Eq.~(\ref{strong2}). The reason for this is that the time averages of the coherences $k_7$ to $k_{10}$ are exactly the same as the quasi-stationary state solutions obtained when setting their time derivatives in Eq.~(\ref{eff}) equal to zero. The solution of Eq.~(\ref{strong2}) implies that the mean phonon number at time $t$ can to a very good approximation be written as
\begin{eqnarray} \label{mfinaldot4}
m (t) &= & \left[ m(0) - m_{\rm ss}  \right] \, {\rm e}^{ - \gamma_{\rm c} t} + m_{\rm ss} 
\end{eqnarray}
in zeroth order in $\eta$ with $m_{\rm ss} = c/\gamma_{\rm c}$, since $n_2$ and $m$ are the same for $\eta \ll 1$. Notice that the general solution in Eq.~(\ref{mfinaldot4}) applies in the strong as well as in the weak confinement regime, although in the weak confinement regime only after a transition time of the order of $1/\alpha_{11}^{(2)}$. In this section, we derive analytical expressions for the stationary state phonon number $m_{\rm ss}$ and for the effective cooling rate $\gamma_{\rm c}$ in this equation which are much more general than the expressions which we obtained already in Section \ref{effeq2}. Afterwards, we show that these rates are in good agreement with numerical solutions of the closed set of 23 cooling equations which can be found in Section \ref{REV} and App.~\ref{appB}. 

\subsection{Stationary state phonon numbers} \label{last}

Using these cooling equations and setting the time derivatives of all expectation values equal to zero, we obtain the stationary state phonon number
\begin{eqnarray} \label{mss}
m_{\rm ss} &=& {1 \over 16 \nu \Delta} \cdot {1 \over \xi_1^4} \, \left[ \xi_2^6 \, \theta - 2 \xi_3^6 \right]
\end{eqnarray}
with the frequencies $\xi_1$, $\xi_2$, and $\xi_3$ defined as
\begin{eqnarray} \label{xi1}
\xi_1^4 &\equiv & (4 \Delta^2 + \Gamma^2) (\Gamma^2 + \nu^2) + 2 (\Gamma^2 + 3 \nu^2) \Omega^2 \, , \notag \\
\xi_2^6 &\equiv & (\Gamma^2 + \nu^2) \left[ (\Gamma^2 + 4 \Delta^2)^2 + 8 (\Gamma^2 - 4 \Delta^2) \nu^2 \right. \notag \\
&& \left. + 16 \nu^4 \right] + 4 \left[ (\Gamma^2 + 2 \nu^2) (\Gamma^2 + 4 \Delta^2) - 8 \nu^4 \right] \Omega^2 \notag \\
&& + 4 (\Gamma^2 + 4 \nu^2) \Omega^4 \, , \notag \\
\xi_3^6 &\equiv & 2 (2 \Delta + \nu) (\Gamma^2 + \nu^2) \left[ \Gamma^2 + 4 (\Delta-\nu)^2 \right] \nu \notag \\
    && + \left[ 3 \Gamma^4 - (4 \Delta^2 - 8 \Delta \nu - 7 \nu^2) \Gamma^2 \right. \notag \\
    && \left. - 4 (\Delta^2 - 6 \Delta \nu + 5 \nu^2) \nu^2 \right] \Omega^2 \, . 
\end{eqnarray}
This result applies in zeroth order in $\eta$ without any approximations and is the main result of this paper. Fig.~\ref{contour} shows $m_{\rm ss}$ as a function of the two laser parameters $\Omega$ and $\Delta$ for different  experimental parameters. In the weak confinement regime one should choose $\Delta = 0.5 \, \Gamma$. However, for $\nu \sim \Gamma$ and $\nu \gg \Gamma$ one should choose $\Delta$ close to $\nu$ in order to minimise the final kinetic energy of the trapped particle.

Fig.~\ref{contour} moreover shows that $m_{\rm ss}$ depends only weakly on the Rabi frequency $\Omega$ which implies that effective laser cooling is not restricted to laser Rabi frequencies much smaller than $\Gamma$ as it is often implied in the literature \cite{Stenholm,sideband2,Stenholm4,ions}. This is an interesting result, since larger $\Omega$'s yield higher cooling rates $\gamma_{\rm c}$, as we shall see below. Our numerical simulations show that $\Omega $ can be as large as $0.3 \, \Gamma$ in the weak confinement and as large as $0.3 \, \nu$ in the strong confinement regime without noticeably increasing the final phonon number $m_{\rm ss}$. However notice that for $\Omega$'s larger than that, cooling changes quickly into heating, since $m_{\rm ss}$ increases relatively rapidly beyond certain critical points.

Let us now have a closer look at different parameter regimes to emphasize that our results are consistent with previous calculations. For example, for relatively small Rabi frequencies $\Omega$, Eq.~(\ref{mss}) simplifies to
\begin{eqnarray} \label{mss2}
m_{\rm ss} &=& {\Gamma^4 + 8 \Gamma^2 (\Delta^2 + \nu^2) + 16 (\Delta^2 - \nu^2)^2 \over 16 \nu \Delta (\Gamma^2 + 4 \Delta^2)} \, \theta \notag \\
&& - { (2 \Delta + \nu) \over 4 \Delta (\Gamma^2 + 4 \Delta^2)} \,  \left[ \Gamma^2 + 4 (\Delta - \nu)^2 \right] \, .
\end{eqnarray}
This equation is in good agreement with the stationary state phonon number implied in Refs.~\cite{Stenholm2,sideband2,Stenholm4,ions} for $d_3=0$. This can be shown using Eqs.~(\ref{apm}) and (\ref{alpha}) in this paper and the expressions for $A_{\pm}$ in Eq.~(7) in Ref.~\cite{Stenholm4}. However, notice that the $A_\pm$ in this paper are slightly different from later definitions of the $A_\pm$ \cite{ions,RevMod}. 

\subsubsection{Weak confinement}

\noindent \begin{figure*}[t]
\begin{minipage}{2\columnwidth}
\begin{center}
\resizebox{\columnwidth}{!}{\rotatebox{0}{\includegraphics{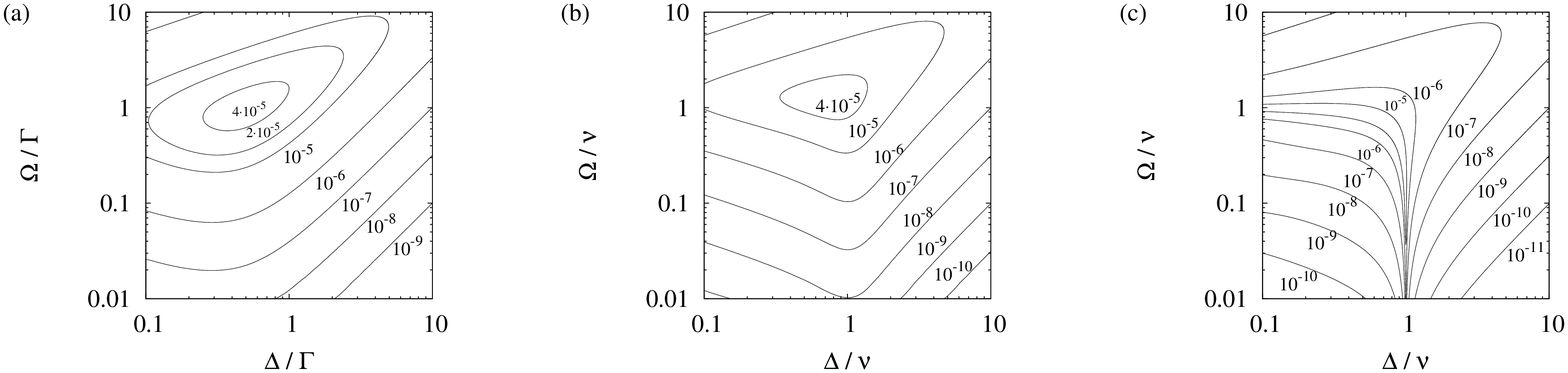}}} 
\end{center}
\caption{Logarithmic contour plot of the cooling rate $\gamma_{\rm c}$ in Eq.~(\ref{gc})  in units of $1/\Gamma$ as a function of the laser parameters $\Omega$ and $\Delta$ for (a) $\nu = 0.01 \, \Gamma$ and $\eta= 0.01$, (b) $\Gamma = \nu$ and $\eta= 0.1$, and (c) $\Gamma = 0.01 \, \nu$ and $\eta= 0.1$.} \label{rates}
\end{minipage}
\end{figure*}

In the weak confinement regime, the stationary state phonon number $m_{\rm ss}$ in Eq.~(\ref{mss2}) simplifies to
\begin{eqnarray} \label{mssweak}
m_{\rm ss}^{\rm weak} &=& {\mu^2 \over 16 \Delta \nu} \, \theta - {(3 \Gamma^2 - 4 \Delta^2) \Omega^2 \over 8 \mu^2 \nu \Delta} \, . 
\end{eqnarray}
Exactly the same stationary state phonon number is obtained when setting the left hand side of the five effective cooling equations in Eq.~(\ref{4888}) equal to zero. This confirms the consistency of the calculations in this paper. As already pointed out above, for small $\Omega$, this expression assumes its minimum if
\begin{eqnarray} \label{oweak}
\Delta &=& {1 \over 2} \Gamma \, .
\end{eqnarray}
For this laser detuning and Rabi frequencies $\Omega \ll \Gamma$, the stationary state phonon number simplifies to 
\begin{eqnarray} \label{mssweak2}
m_{\rm ss}^{\rm weak} &=& {\Gamma \over 4 \nu} \, \theta 
\end{eqnarray}
which is much larger than one.

\subsubsection{Strong confinement}

Using the effective cooling equation derived in Section \ref{effeq2}, we find that the stationary state phonon number in the strong confinement regimes equals
\begin{eqnarray} \label{more}
m_{\rm ss}^{\rm strong} &=& {(\Delta - \nu)^2 \over 4 \nu \Delta^3} \, \left[ (\Delta + \nu)^2 \, \theta  - (2 \Delta + \nu) \nu \right] ~~~
\end{eqnarray}
to a very good approximation. Exactly the same result is obtained when neglecting terms proportional to $\Gamma$ and $\Omega^2$ in Eq.~(\ref{mss}). This result suggests immediately that one should choose 
\begin{eqnarray} \label{ostrong}
\Delta &=& \nu 
\end{eqnarray}
in order to minimise the final phonon number $m_{\rm ss}$. When substituting this detuning into Eq.~(\ref{mss2}), we find that the stationary state phonon number for laser sideband cooling is for relatively small Rabi frequencies $\Omega$ to a very good approximation given by
\begin{eqnarray} \label{mssstrong}
m_{\rm ss} &=& {\Gamma^2 \over 16 \nu^2} \, \left[ 4 \theta - 3 \right] 
\end{eqnarray}
which is much smaller than one \cite{sideband}.

\subsection{Effective cooling rates} \label{effrates}

Proceeding as above but eliminating only the expectation values other than $n_2$ adiabatically from the time evolution of the trapped particle, we find that the effective cooling rate $\gamma_{\rm c}$ in Eq.~(\ref{mfinaldot4}) is in second order in $\eta$ given by  
\begin{eqnarray} \label{gc}
\gamma_{\rm c} &=& {16 \eta^2 \nu \Delta \Gamma \Omega^2 \over \mu^2} \cdot {\xi_1^4 \over \xi_2^6}
\end{eqnarray}
with $\xi_1$ and $\xi_2$ as in Eq.~(\ref{xi1}). Fig.~\ref{rates} shows this cooling rate as a function of the laser parameters $\Omega$ and $\Delta$. Indeed we find that $\gamma_{c}$ increases in general as $\Omega$ increases. This confirms that one should choose $\Omega$ in general as large as possible without noticeably decreasing the stationary state phonon number $m_{\rm ss}$.

\noindent \begin{figure*}[t]
\begin{minipage}{2\columnwidth}
\begin{center}
\resizebox{\columnwidth}{!}{\rotatebox{0}{\includegraphics{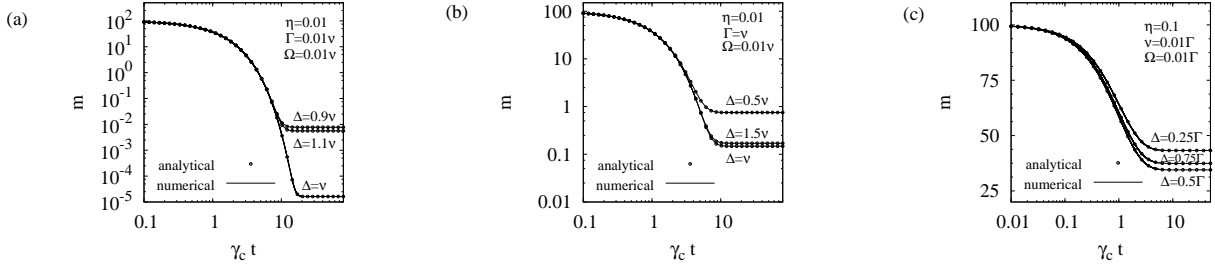}}} 
\end{center}
\caption{Logarithmic plot of the time dependence of the mean phonon number $m$ for the experimental parameters indicated in the figure. Here (a) illustrates the strong confinement, (b) shows the medium confinement, and (c) shows the weak confinement regime. The numerical solutions are the result of a numerical integration of 23 cooling equations while the analytical solution represents Eq.~(\ref{mfinaldot4}).} \label{ions1}
\end{minipage}
\end{figure*}

For relatively small Rabi frequencies $\Omega$, i.e.~for $\Omega \ll \Gamma$, the cooling rate $\gamma_{\rm c}$ in Eq.~(\ref{gc}) simplifies to
\begin{eqnarray}
\gamma_{\rm c} &=& {\eta^2 \Gamma \Omega^2 \over \Gamma^2 + 4 (\Delta - \nu)^2} - {\eta^2 \Gamma \Omega^2 \over \Gamma^2 + 4 (\Delta + \nu)^2} \, . 
\end{eqnarray}
When comparing this result with the expression for $A_- - A_+$ in \cite{Stenholm2,Stenholm4}, we find that 
our cooling rate is essentially the same as the cooling rate reported in these earlier references. Moreover, as Fig.~\ref{ions1} confirms, Eq.~(\ref{gc}) is in very good agreement with numerical solutions. 

\subsubsection{Weak confinement}

In the weak confinement regime, the laser detuning $\Delta $ which minimises the stationary state phonon number $m_{\rm ss}$ is given by ${1 \over 2} \, \Gamma$ (cf.~Eq.~(\ref{oweak})). Taking this into account, the cooling rate $\gamma_{\rm c}$ in Eq.~(\ref{gc}) simplifies to
\begin{eqnarray}
\gamma_{\rm c} &=& {2 \eta^2 \nu \Omega^2 \over \Gamma^2} 
\end{eqnarray}
in the limit of weak driving, ie.~when neglecting terms proportional to $\Omega^2$. As Fig.~\ref{ions1}(c) illustrates, this cooling rate is in good agreement with a numerical solution of the full set of 23 rate equations.

\subsubsection{Strong confinement}

In the strong confinement regime, terms which scale as $\Gamma$ or $\Omega$ are in general negligible (cf.~Eq.~(\ref{cond})). Taking this into account and simplifying Eq.~(\ref{gc}) accordingly, we find that the cooling rate $\gamma_{\rm c}$ in this equation is exactly the same as $\gamma_{\rm c}$ in Eq.~(\ref{eff88strong}). The cooling process becomes indeed the most efficient, when the detuning $\Delta$ is close to the phonon frequency $\nu$ (cf.~Eq.~(\ref{ostrong})). The cooling rate is in this case given by
\begin{eqnarray}
\gamma_{\rm c} &=& {\eta^2 \Omega^2 \over \Gamma} 
\end{eqnarray}
which is essentially the stationary state photon emission rate of a laser-driven atom multiplied by $\eta^2$. 

\section{Conclusions} \label{conc}

This paper analyses the cooling process of a single trapped particle with red-detuned laser light. In contrast to previous authors \cite{Stenholm2,Stenholm4,sideband2,ions,RevMod}, our analysis avoids the adiabatic elimination of the excited atomic state when calculating the effective cooling rate $\gamma_{\rm c}$ and the stationary state phonon number $m_{\rm ss}$. Our calculations hence apply to a wider range of laser Rabi frequencies $\Omega$. They show that $\Omega$ can be chosen relatively large without affecting the final outcome of the cooling process. As illustrated in Fig.~\ref{contour}, in the weak confinement regime with $\nu \ll \Gamma$, the Rabi frequency $\Omega $ can be as large as $0.3 \, \Gamma$ and in the strong confinement regime with $\Gamma \ll \nu$, the Rabi frequency $\Omega $ can be as large as $0.3 \, \nu$ without reducing the stationary state phonon number $m_{\rm ss}$ noticeably. This is an interesting observation, since $\gamma_{\rm c}$ scales as $\Omega^2$ and increases rapidly when $\Omega$ increases.

The main novelty of our calculations is a transformation of the original Hamiltonian which replaces the atomic lowering operator $\sigma^-$ and the phonon annihilation operator $b$ by two new operators $x$ and $y$ which commute with each other. The corresponding particles are neither atomic excitations nor phonons but provide a more natural description of the trapped particle. Our calculations are therefore more straightforward than previous calculations. Since the theory of laser cooling has already been studied in great detail in the literature~\cite{Stenholm2,key,RevMod,ions}, the main purpose of this paper is to establish and test a framework for the modeling of laser cooling which can be extended relatively easily to more complex cooling scenarios like cavity-mediated laser cooling \cite{cavity,cavity2} and possible quantum optical heating mechanisms in sonoluminescence experiments \cite{SL}. \\[0.5cm]
 
\noindent {\em Acknowledgement.} This work was supported by the UK Research Council EPSRC. N. S. S. acknowledges funding from The Ministry of Education in Saudi Arabia.

\appendix
\section{Derivation of the reset state ${\cal R}(\rho)$} \label{app}

Substituting the interaction Hamiltonian $H_{\rm I}$ in Eq.~(\ref{2.8}) into the expression for ${\cal R}(\rho)$ in Eq.~(\ref{long}) and using first order perturbation theory we find that 
\begin{eqnarray}
{\cal R} (\rho) &=& \lim_{\Delta t \to 0} {1 \over \Delta t} \int_0^{\Delta t} {\rm d} t \int_0^{\Delta t} {\rm d}t' \sum_{{\bf k} \lambda} \notag \\
&& \langle 1_{{\bf k} \lambda}| \, H_{\rm I} (t)  |0 \rangle \, \rho \,  \langle 0| \, H_{\rm I} (t') \, |1_{{\bf k} \lambda} \rangle \, .
\end{eqnarray}
Using the concrete expression for $H_{\rm I} (t)$ in Eq.~(\ref{2.8}), the relation
\begin{eqnarray}
\int_0^{\Delta t} {\rm d} t' \, {\rm e}^{{\rm i}(\omega_k - \omega_{\rm L})(t-t')} = 2 \pi \delta(\omega_k - \omega_{\rm L}) \, , 
\end{eqnarray}
and ignoring an overall level shift which can be absorbed into the free energy of the system, the reset operator ${\cal R} (\rho) $ becomes
\begin{eqnarray}
{\cal R} (\rho) &=& \sum_{{\bf k} \lambda} 2 \pi \, |g_{{\bf k} \lambda}|^2 \, \delta(\omega_k - \omega_{\rm L}) \nonumber \\
&& \sigma^{-} D \left({{\rm i} \eta k \cos \vartheta \over k_{\rm L}} \right) \, \rho \, D \left(- {{\rm i} \eta k \cos \vartheta \over k_{\rm L}} \right) \sigma^+ \, . ~~~~~
\end{eqnarray}
This expression can be evaluated relatively easily in the large volume limit, where
\begin{eqnarray}
\sum_{{\bf k} \lambda} &\longleftarrow & \sum_{\lambda=1,2} \left( {L \over 2\pi c} \right)^3 \int_0^\infty {\rm d} \omega_k \, \omega_k^2 \int_0^\pi {\rm d} \vartheta \, \sin \vartheta \int_0^{2 \pi} {\rm d} \varphi \, , \nonumber \\
\end{eqnarray}
when using the same notation as in Section \ref{symi}. Performing the integration implied by this equation, using Eqs.~(\ref{2.4}) and (\ref{symi2}), and writing the normalised dipole moment $\hat {\bf D}_{01} \equiv {\bf D}_{01}/|{\bf D}_{01}|$ as
\begin{eqnarray}
\hat {\bf D}_{01} &=& (d_1,d_2,d_3)^{\rm T} 
\end{eqnarray}
we finally find that ${\cal R} (\rho) $ is indeed given by Eq.~(\ref{longfinal}). The spontaneous decay rate $\Gamma$ in this equation is the same as $\Gamma $ in Eq.~(\ref{symi3}). Notice that $d_3$ denotes the component of the normalised dipole moment $\hat {\bf D}_{01}$ in the direction of the cooling laser.

\section{Relevant expectation values} \label{appA}

The calculations in the following two appendices require in addition to the expectation values defined in Section \ref{REV} the $x$ operator expectation values
\begin{eqnarray} \label{coherencesappA}
&& \hspace*{-0.4cm} k_1 \equiv \langle x + x^\dagger \rangle \, , ~~
k_2 \equiv {\rm i} \, \langle x - x^\dagger \rangle \, .
\end{eqnarray}
Moreover we employ the mixed operator expectation values 
\begin{eqnarray} \label{coherences3}
&& \hspace*{-0.3cm} n_4 \equiv \langle x^\dagger x y^\dagger y \rangle \, , ~~ 
k_{13} \equiv \langle (x + x^\dagger ) y^\dagger y \rangle \, , \notag \\
&& \hspace*{-0.3cm} k_{14} \equiv {\rm i}  \, \langle (x - x^\dagger ) y^\dagger y \rangle \, , ~~
k_{15} \equiv \langle (x - x^\dagger ) (y - y^\dagger ) \rangle \, , \notag \\
&& \hspace*{-0.3cm} k_{16} \equiv {\rm i} \, \langle(x + x^\dagger ) (y - y^\dagger ) \rangle \, , ~~
k_{17} \equiv \langle (x + x^\dagger ) (y + y^\dagger ) \rangle \, , \notag \\
&& \hspace*{-0.3cm} k_{18} \equiv {\rm i} \, \langle(x - x^\dagger ) (y + y^\dagger ) \rangle \, , 
\end{eqnarray}
and
\begin{eqnarray} \label{coherences3}
k_{19} &\equiv & \langle (x - x^\dagger) (y^2 - y^{\dagger 2}) \rangle \, , \nonumber \\ 
k_{20} &\equiv & {\rm i} \, \langle (x + x^\dagger) (y^2 - y^{\dagger 2} ) \rangle \, , \nonumber \\ 
k_{21} &\equiv & \langle (x + x^\dagger) (y^2 + y^{\dagger 2} ) \rangle \, , \nonumber \\ 
k_{22} &\equiv & {\rm i} \, \langle (x - x^\dagger) (y^2 + y^{\dagger 2} ) \rangle \, , \notag \\
k_{23} &\equiv & \langle x^\dagger x (y^2 + y^{\dagger 2} ) \rangle \, , \nonumber \\ 
k_{24} &\equiv & {\rm i} \, \langle x^\dagger x (y^2 - y^{\dagger 2} ) \rangle \, .   
\end{eqnarray}
The time derivatives of these and other expectation values which we defined in Sec.~\ref{REV} can be found in App.~\ref{appB}. Notice that this paper does not introduce expectation values $n_3$ and $k_3$ to $k_6$. These names are reserved for variables which are used in other cooling scenarios.

\section{$n_1$, $k_{11}$, and $k_{12}$ in the weak confinement regime} \label{appB}

Setting $\eta = 0$ and substituting the $x$ operator expectation values $n_1$, $k_1$, and $k_2$ into Eq.~(\ref{dotA}), we find that they evolve according to
\begin{eqnarray} \label{48882}
\dot n_1 &=& {1 \over 2} \Omega \, k_2 - \Gamma \, n_1  \, , \nonumber \\
\dot k_1 &=& - \Delta \, k_2 - {1 \over 2} \Gamma \, k_1 \, , \nonumber \\ 
\dot k_2 &=& \Omega (1 - 2 n_1) + \Delta \, k_1 -  {1 \over 2} \Gamma \, k_2 
\end{eqnarray}
in zeroth order in $\eta$. These equations form a closed set of differential equations. Eliminating all $x$-operator expectation values adiabatically which change on the relatively fast time scale given by $\Gamma$ and adopting a notation where $x= x^{(0)} + x^{(1)} + x^{(2)} + ... $, with the superscript indicating the scaling of the respective term with respect to $\eta$, we find for example that $n_1$ is in zeroth order in $\eta$ given by
\begin{eqnarray} \label{zeroth}
n_1^{(0)} &=& {\Omega^2 \over \mu^2} \, . ~~
\end{eqnarray}
The constant $\mu^2$ in this equation is given by
\begin{eqnarray} \label{eff44}
\mu^2 &\equiv & 2 \Omega^2 + \Gamma^2 + 4 \Delta^2 \, .
\end{eqnarray}
In addition to $n^{(0)}$, we obtain solutions for $k_1^{(0)}$ and $k_2^{(0)}$. These will be used in the next subsection to calculate the coherences $k_{11}$ and $k_{12}$ up to first order in $\eta$.

Setting $\eta = 0$ and using again Eq.~(\ref{dotA}), we find that the time evolution of the mixed operator coherences $k_{11}$ and $k_{12}$ and $k_{15}$ to $k_{18}$ is in zeroth order in $\eta$ given by
\begin{eqnarray} \label{7772}
\dot k_{11} &=& {1 \over 2} \Omega \, k_{18} - \nu \, k_{12} - \Gamma \, k_{11} \, , \nonumber \\ 
\dot k_{12} &=& - {1 \over 2} \Omega \, k_{15} + \nu \, k_{11} - \Gamma \, k_{12} \, , \nonumber \\
\dot k_{15} &=& - \Omega (k_8 - 2 k_{12}) - \Delta \, k_{16} - \nu \, k_{18} - {1 \over 2} \Gamma \, k_{15} \, , ~~\nonumber \\
\dot k_{16} &=& \Delta \, k_{15} + \nu \, k_{17} - {1 \over 2} \Gamma \, k_{16} \, , \nonumber \\
\dot k_{17} &=& - \Delta \, k_{18} - \nu \, k_{16} - {1 \over 2} \Gamma \, k_{17} \, , \nonumber \\
\dot k_{18} &=& \Omega (k_7 - 2 k_{11} ) + \Delta \, k_{17} + \nu \, k_{15} - {1 \over 2} \Gamma \, k_{18} \, .
\end{eqnarray}
All six expectation values evolve on the relatively fast time scale given by the spontaneous decay rate $\Gamma$. Taking this into account and eliminating them adiabatically in the weak coupling regime, i.e.~for relatively small $\nu$, we find that
\begin{eqnarray} \label{k11120}
k_{11}^{(0)} &=& {\Omega^2 \over \mu^4 \Gamma} \left[ \mu^2 \Gamma \, k_7 - (3 \Gamma^2- 4 \Delta^2) \nu \, k_8 \right] \, , \notag\\
k_{12}^{(0)} &=& {\Omega^2 \over \mu^4 \Gamma} \left[ (3 \Gamma^2- 4 \Delta^2) \nu  \, k_7 + \mu^2 \Gamma  \, k_8 \right] 
\end{eqnarray}
to a very good approximation. The constant $\mu^2$ is given in Eq.~(\ref{eff44}) above. In addition to $k_{11}^{(0)}$ and $k_{12}^{(0)}$ we obtain expressions for $k_{15}^{(0)}$ and $k_{16}^{(0)}$. These are used in the next subsection to calculate $n_1$ up to first order in $\eta$.

Proceeding as above but taking terms up to first order in $\eta$ into account we find that the first order in $\eta$ contributions of the $x$ operator expectation values $n_1$, $k_1$, and $k_2$ in Eq.~(\ref{coherences}) evolve according to
\begin{eqnarray} \label{48882}
\dot n_1^{(1)} &=& {1 \over 2} \Omega \, k_2^{(1)} - \Gamma \, n_1^{(1)}  \, , \nonumber \\
\dot k_1^{(1)} &=& - \eta \nu \, k_{15}^{(0)} - \Delta \, k_2^{(1)} - {1 \over 2} \Gamma \, k_1^{(1)} \, , \nonumber \\ 
\dot k_2^{(1)} &=& - 2 \Omega n_1^{(1)} - \eta \nu \, k_{16}^{(0)} + \Delta \, k_1^{(1)} -  {1 \over 2} \Gamma \, k_2^{(1)} \, . ~~~~
\end{eqnarray}
These equations form a closed set of cooling equations, when the results for $k_{15}^{(0)}$ and $k_{16}^{(0)}$ which we obtained in App.~\ref{appB} are taken into account. Eliminating $n_1$, $k_1$ and $k_2$ adiabatically
\begin{eqnarray} \label{zeroth2}
n_1^{(1)} &=& {8 \eta \nu \Delta \Omega^2 \over \mu^4} \, k_8  
\end{eqnarray}
in the weak confinement regime which we introduced in Section \ref{effeq}. This means terms proportional to $\nu^2$ have been neglected.

\begin{figure}[t]
\begin{minipage}{\columnwidth}
\begin{center}
\hspace*{-3cm} \includegraphics[scale=0.9]{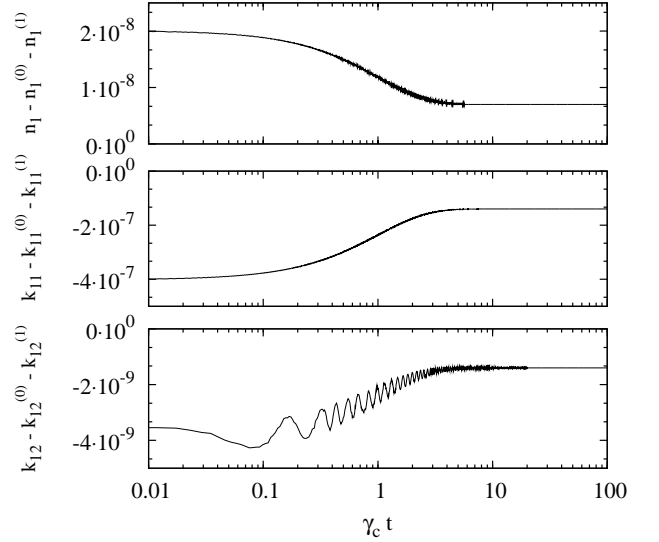}
\end{center}
\caption{A comparison of the analytical results for $n_1$, $k_{11}$, and $k_{12}$ in Eqs.~(\ref{zeroth}), (\ref{k11120}), (\ref{zeroth2}), and (\ref{k11121}) with the results of a numerical solution of the above cooling equations for $\eta = 0.1$, $\Omega = \nu = 0.01 \, \Gamma$, and $\Delta = 0.5 \, \Gamma$.} \label{ion1}
\end{minipage}
\end{figure}

In order to calculate $k_{11}$ and $k_{12}$ up to first order in $\eta$, we need a closed set of cooling equations which holds correctly up to this order. Applying Eq.~(\ref{dotA}) again to $k_{11}$ and $k_{12}$ and $k_{15}$ to $k_{18}$, we find that 
\begin{eqnarray} \label{777}
\dot k_{11}^{(1)} &=& {1 \over 2} \Omega \, k_{18}^{(1)} - \nu \, k_{12}^{(1)} + 2 \eta \nu \, n_1^{(0)} - \Gamma \, k_{11}^{(1)} \, , \nonumber \\ 
\dot k_{12}^{(1)} &=& - {1 \over 2} \Omega \, k_{15}^{(1)} + \nu \, k_{11}^{(1)} - \Gamma \, k_{12}^{(1)} \, , \nonumber \\
\dot k_{15}^{(1)} &=& 2 \Omega \, k_{12}^{(1)} - \Delta \, k_{16}^{(1)} - \nu \, k_{18}^{(1)} + \eta \nu \left( k_1^{(0)} + 2 k_{13}^{(0)} \right . \nonumber \\
&& \left. - k_{21}^{(0)} \right) - {1 \over 2} \Gamma \, k_{15}^{(1)} \, , \nonumber \\
\dot k_{16}^{(1)} &=& \Delta \, k_{15}^{(1)} + \nu \, k_{17}^{(1)} + \eta \nu \left( k_2^{(0)} + 2 k_{14}^{(0)} - k_{22}^{(0)} \right) \nonumber \\
&& - {1 \over 2} \Gamma \, k_{16}^{(1)}  \, , \nonumber \\
\dot k_{17}^{(1)} &=& - \Delta \, k_{18}^{(1)} - \nu \, k_{16}^{(1)} + \eta \nu \left( k_1^{(0)} - k_{19}^{(0)} \right) - {1 \over 2} \Gamma \, k_{17}^{(1)}  \, , \nonumber \\
\dot k_{18}^{(1)} &=& - 2 \Omega \, k_{11}^{(1)} + \Delta \, k_{17}^{(1)} + \nu \, k_{15}^{(1)} + \eta \nu \left( k_2^{(0)} - k_{20}^{(0)} \right) \notag \\
&& - {1 \over 2} \Gamma \, k_{18}^{(1)} \, . 
\end{eqnarray}
Substituting the definitions of the mixed-particle expectation values $n_4$, $k_{13}$ and $k_{14}$ and $k_{19}$ to $k_{22}$ into Eq.~(\ref{dotA}) and setting $\eta = 0$, we find that 
\begin{eqnarray} \label{48222}
\dot n_4 &=&{1 \over 2} \Omega \, k_{14} - \Gamma \, n_4 \, , \notag \\
\dot k_{13} &=& - \Delta \, k_{14} - {1 \over 2} \Gamma \, k_{13} \, , \nonumber \\ 
\dot k_{14} &=& \Omega \, (n_2 - 2 n_4) + \Delta \, k_{13} - {1 \over 2} \Gamma \, k_{14} \, , 
\end{eqnarray}
while
\begin{eqnarray} \label{48222x}
\dot k_{19} &=& - \Omega ( k_{10} - 2 k_{24}) - \Delta \, k_{20} - 2 \nu \, k_{22} -  {1 \over 2} \Gamma \, k_{19} \, , \nonumber \\ 
\dot k_{20} &=& \Delta \, k_{19} + 2 \nu \, k_{21} - {1 \over 2} \Gamma \, k_{20} \, , \nonumber \\
\dot k_{21} &=& - \Delta \, k_{22} - 2 \nu \, k_{20} -  {1 \over 2} \Gamma \, k_{21} \, , \nonumber \\ 
\dot k_{22} &=& \Omega ( k_9 - 2 k_{23} )+ \Delta \, k_{21} + 2 \nu \, k_{19} -  {1 \over 2} \Gamma \, k_{22} \notag \\
\dot k_{23} &=& {1 \over 2} \Omega \, k_{22} - 2 \nu \, k_{24} - \Gamma \,  k_{23} \, , \nonumber\\
\dot k_{24}&=& - {1\over 2} \Omega \, k_{19} + 2 \nu \, k_{23} - \Gamma \, k_{24} \, .
\end{eqnarray}
These final six differential equations hold in zeroth order in $\eta$. Setting the right hand side of these and of the cooling equations in Eq.~(\ref{777}) equal to zero, we finally find that
\begin{eqnarray} \label{k11121}
k_{11}^{(1)} &=& {4 \eta \nu \Omega^2 \over \mu^4} \, \left[ 2 \Delta \, k_{10} + \Gamma \right] \, , \notag \\
k_{12}^{(1)} &=& {8 \eta \nu \Delta \Omega^2 \over \mu^4} \, \left[ 2 n_2 - k_9 + 1 \right] 
\end{eqnarray}
in the weak confinement regime. This means, terms of order $\nu^2$ have again been neglected. Fig.~\ref{ion1} compares the above analytical results for $n_1$, $k_{11}$, and $k_{12}$ with the result of a numerical solution of the above cooling equations. Very good agreement between both solutions is found which suggests that the effective cooling equations in Eq.~(\ref{eff}) apply after a very short transition time of the order $1/\Gamma$.

\section{$n_1$, $k_{11}$ and $k_{12}$ in the strong confinement regime} \label{appF}

\begin{figure}[t]
\begin{minipage}{\columnwidth}
\begin{center}
\hspace*{-3cm} \includegraphics[scale=0.9]{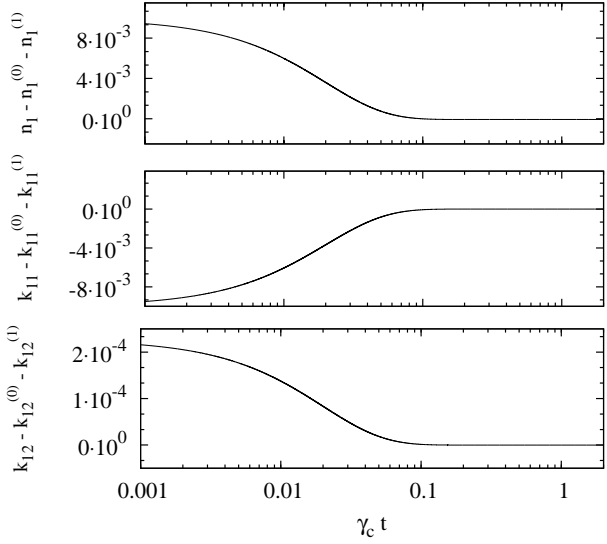}
\end{center}
\caption{A comparison of the analytical results for $n_1$, $k_{11}$, and $k_{12}$ in Eqs.~(\ref{zerothxxx}), (\ref{xxx}), and (\ref{k11121xxx}) with the results of a numerical solution of the above cooling equations for $\eta = 0.01$, $\Omega = \Gamma = 0.01 \, \nu$, and $\Delta = \nu$.} \label{ion2}
\end{minipage}
\end{figure}

The calculation of $n_1$ in zeroth order in $\eta$ is the same as in App.~\ref{appB}. However, in the strong confinement regime, the expression in Eq.~(\ref{zeroth}) simplifies to
\begin{eqnarray} \label{zerothxxx}
n_1^{(0)} &=& {\Omega^2 \over 4 \Delta^2} \, . ~~
\end{eqnarray}
Setting $\eta = 0$ and eliminating the $y$ operator coherences adiabatically from the system dynamics we immediately find that $k_7$ to $k_{10}$ all equal zero in zeroth order in $\eta$,
\begin{eqnarray} \label{nice}
k_7^{(0)} = k_8^{(0)} = k_9^{(0)} = k_{10}^{(0)} =0 \, .
\end{eqnarray}
Taking this into account when eliminating the mixed operator expectation values whose time derivatives are given in Eq.~(\ref{7772}), we moreover find that 
\begin{eqnarray} \label{xxx}
k_{11}^{(0)} = k_{12}^{(0)} =0 \, .
\end{eqnarray}
To calculate the coherences $k_{11}$ and $k_{12}$ up to first order in $\eta$, we have a look at the time derivatives of  $k_{11}$, $k_{12}$, and $k_{15}$ to $k_{18}$ in first order in $\eta$ which can be found in Eq.~(\ref{777}). Combining Eqs.~(\ref{48222x}) and (\ref{nice}), we immediately see that
\begin{eqnarray} \label{nice2}
k_{19}^{(0)} = k_{20}^{(0)} = k_{21}^{(0)} = k_{22}^{(0)} =0 \, .
\end{eqnarray}
Taking this and the expressions for $k_1^{(0)}$, $k_2^{(0)}$, $k_{13}^{(0)}$, and $k_{14}^{(0)}$ obtained in App.~\ref{appB} into account, when setting the time derivatives of the relatively fast evolving variables in Eq.~(\ref{777}) equal to zero, we therefore find that  
\begin{eqnarray} \label{k11121xxx}
k_{11}^{(1)} &=& {\eta \nu \Gamma \Omega^2 \over 4 (\Delta + \nu)^2 \Delta^2} \, \left[ 1 - {4 \nu \Delta \over (\Delta - \nu)^2} \, n_2  \right] \, , \nonumber \\
k_{12}^{(1)} &=& {\eta \nu \Omega^2 \over 2 (\Delta + \nu) \Delta^2}  \left[ 1 + {2 \Delta \over \Delta - \nu} \, n_2 \right] \, .
\end{eqnarray}
These coherences are different from the coherences in Eq.~(\ref{k11121}), since they apply only in the strong confinement regime. As Fig.~\ref{ion2} shows there is again very good agreement between the analytical and the numerical solutions for $n_1$, $k_{11}$, and $k_{12}$. This means that the effective cooling equation for the strong confinement regime in Eq.~(\ref{strong2}) too applies after a relatively short transition time.

\end{document}